\documentclass[aps,pra,superscriptaddress,twocolumn,10pt]{revtex4-2}

\usepackage{amsmath}
\usepackage{amssymb}
\usepackage{amsthm}
\usepackage{bbm}
\usepackage{color}
\usepackage{graphicx}[floatfix]
\usepackage{hyperref}
\usepackage[utf8]{inputenc}
\usepackage[T1]{fontenc}
\usepackage{physics}
\usepackage{times}

\hypersetup{
    colorlinks=true,linkcolor=blue,citecolor=blue,
    filecolor=blue,urlcolor=blue,breaklinks=true
}

\newcommand{\orcid}[1]{\href{https://orcid.org/#1}
{\includegraphics[width=7pt]{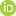}}}

\theoremstyle{definition}

\def\be{\begin{equation}}
\def\ee{\end{equation}}

\def\bc{\begin{center}}
\def\ec{\end{center}}
\def\bal{\begin{align}}
\def\eal{\end{align}}

\def\ie{i.e.}
\def\etal{\textit{et al.}}

\begin{document}

\title{Effects of quantum deformation on the Jaynes-Cummings and
  anti-Jaynes-Cummings models}

\author{Gustavo M. Uhdre\orcid{0000-0003-0344-8532}}
\affiliation{
  Programa de P\'os-Gradua\c{c}\~{a}o em Ci\^{e}ncias/F\'{i}sica,
  Universidade Estadual de Ponta Grossa,
  84030-900 Ponta Grossa, Paran\'a, Brazil
}

\author{Danilo Cius\orcid{0000-0002-4177-1237}}
\affiliation{
  Programa de P\'os-Gradua\c{c}\~{a}o em Ci\^{e}ncias/F\'{i}sica,
  Universidade Estadual de Ponta Grossa,
  84030-900 Ponta Grossa, Paran\'a, Brazil
}

\author{Fabiano M. Andrade\orcid{0000-0001-5383-6168}}
\email{fmandrade@uepg.br}
\affiliation{
  Programa de P\'os-Gradua\c{c}\~{a}o em Ci\^{e}ncias/F\'{i}sica,
  Universidade Estadual de Ponta Grossa,
  84030-900 Ponta Grossa, Paran\'a, Brazil
}
\affiliation{
  Departamento de Matem\'{a}tica e Estat\'{i}stica,
  Universidade Estadual de Ponta Grossa,
  84030-900 Ponta Grossa, Paran\'{a}, Brazil
}

\date{\today}

\begin{abstract}
The theory of non-Hermitian systems and the theory of quantum
deformations have attracted a great deal of attention in the past
decades.
In general, non-Hermitian Hamiltonians are constructed by an
\textit{ad hoc} manner.
Here, we study the (2+1) Dirac oscillator and show that in the context
of the $\kappa$--deformed Poincar\'e-Hopf algebra its Hamiltonian is
non-Hermitian but has real eigenvalues.
The non-Hermiticity stems from the $\kappa$-deformed algebra.
From the mapping in
Bermudez \etal, \href{https://doi.org/10.1103/PhysRevA.76.041801}{Phys. Rev. A \textbf{76}, 041801(R) (2007)},
we propose the $\kappa$-Jaynes-Cummings and
$\kappa$-anti-Jaynes-Cummings models, which describe an
interaction between a two-level system with a quantized mode of an
optical cavity in the $\kappa$-deformed context.
We find that the $\kappa$-deformation modifies the
\textit{Zitterbewegung} frequencies and the collapses and
revivals of quantum oscillations.
In particular, the total angular momentum in the $z$ direction is not
conserved anymore, as a direct consequence of the deformation.\\

\noindent DOI: \href{https://doi.org/10.1103/PhysRevA.105.013703}{10.1103/PhysRevA.105.013703}
\end{abstract}


\maketitle

\section{Introduction}
\label{sec:introduction}

The interest in non-Hermitian Hamiltonians with real spectrum
started with the seminal work of Bender and Boettcher
\cite{PRL.80.5243.1998}.
In the past two decades, these systems have been discussed in connection
with invariance under spatio-temporal reflection.
A $\mathcal{PT}$-symmetric Hamiltonian is invariant under spatial
reflection ($\mathcal{P}$) and time-reversal ($\mathcal{T}$) symmetries \cite{CP.46.277.2005,RPP.70.947.2007}.
Many applications of $\mathcal{PT}$-symmetric Hamiltonians are found in
the study of gain and loss systems \cite{PRB.98.125102.2018} which may
be found in different physical contexts \cite{NPhot.13.883.2019}.
In standard quantum mechanics, the Hermiticity, or being more precise,
the self-adjointness of physical observables, especially of
Hamiltonians, guarantees that the quantum evolution is unitary and the
spectrum is real.
If the eigenstates of the Hamiltonian and the $\mathcal{PT}$ operator
are the same, it is said to have an unbroken $\mathcal{PT}$ symmetry,
and the $\mathcal{PT}$-symmetric Hamiltonian is also quasi-Hermitian
\cite{AoP.213.74.1992,PRA.91.052113.2015}.
From the theory of quasi-Hermitian operators, we know that it has real
eigenvalues but the time evolution is not unitary.
However, for time-independent non-Hermitian Hamiltonians
\cite{PRA.93.042114.2016}, it is possible to have a unitary evolution if
we employ a similarity transformation \cite{PR.102.1230.1956} which
leads to its Hermitian counterpart.

In parallel and separately, in the past decades the theory of quantum
deformations based on the $\kappa$-Poincar\'{e}-Hopf algebra has also
attracted a great deal of attention and has been an alternative
framework for studying relativistic and non-relativistic quantum systems
and represents an interesting theory due to its phenomenological
applications.
The $\kappa$-deformed Poincar\'{e}-Hopf algebra, established
in Refs. \cite{PLB.264.331.1991,PLB.293.344.1992,PLB.329.189.1994,
PLB.334.348.1994}, is based on the following commutation relations
\begin{subequations}
\label{eq:algebra}
\begin{align}
\left[P_{\nu},P_{\mu}\right]= {} & 0,\\
\left[M_{i},P_{\mu}\right] = {} &
(1-\delta_{0\mu})i\epsilon_{ijk}P_{k},\\
\left[L_{i},P_{\mu}\right]= {} & i [P_{i}]^{\delta_{0\mu}}
[\delta_{ij}\varepsilon^{-1}
\sinh \left( \varepsilon P_{0}\right)]^{1-\delta_{0\mu}},\\
\left[M_{i},M_{j}\right]= {} & i\epsilon_{ijk} M_{k},\qquad
\left[M_{i},L_{j}\right]=i\epsilon_{ijk} L_{k},\\
  \left[L_{i},L_{j}\right]= {} &
  -i\epsilon_{ijk}
  \left[
    M_{k}\cosh \left(\varepsilon P_{0}\right)-
    \frac{\varepsilon ^{2}}{4}P_{k}P_{l} M_{l}
  \right],
\end{align}
\end{subequations}
where $\varepsilon$ is defined by
\begin{equation}
  \varepsilon=\frac{1}{\kappa}=
  \lim_{R\rightarrow \infty }(R\ln q),
\end{equation}
with $R$ being the de Sitter curvature and $q$ a real
deformation parameter, $P_{\mu }=(P_{0},\boldsymbol{P})$
are the $\kappa$-deformed generators for energy and momenta, and
$M_{i}$ and $L_{i}$ represent the spatial rotations and deformed boost
generators, respectively.
The parameter $\kappa$ has the dimension of mass and claimed from the
very beginning that it must have something to do with quantum gravity,
and therefore it is usually interpreted as being the Planck mass $M_{P}$
\cite{PLB.711.122.2012}.
We also comment that in  Ref. \cite{EPJC.78.665.2018}, it was discussed
that if the parameter $\kappa$ does not correspond to an observable,
then its value should be inferred through some indirect measurements.
For a short introduction to the $\kappa$-deformation framework,
see Ref. \cite{IJMPA.32.1730026.2017}.
In the context of $\kappa$-deformed theory, the physical properties of
relativistic quantum mechanics can be addressed by solving the
$\kappa$-deformed Dirac equation
\cite{PLB.302.419.1993,PLB.318.613.1993,CQG.21.2179.2004,JHEP.2004.28.2004}.
For instance, it has implications in the divergenceless of the vacuum
energy in quantum field theory \cite{PRD.76.125005.2007} and
in the spin-1/2 Aharonov-Bohm problem \cite{PLB.719.467.2013} leading to
additional bound states \cite{PLB.359.339.1995}, as well as in the
in the Landau levels \cite{PLB.339.87.1994,EPL.116.31002.2016} and in
the two-dimensional (2D) and three-dimensional (3D) Dirac oscillators \cite{PLB.731.327.2014,PLB.738.44.2014}.

As stated above, although some quantum systems could be effectively
described by non-Hermitian Hamiltonians as considered, for instance,
in Refs. \cite{PRX.4.041001.2014,NPhys.15.1232.2019,S.371.1240.2021},
non-Hermitian systems are usually constructed by
exactly balancing loss and gain \cite{PRL.100.103904.2008} and this
is usually achieved in an \textit{ad hoc} manner.
In the present work, we revisit the 2D Dirac oscillator, the
relativistic version of the simple harmonic oscillator (see below) and
show that in the context of the $\kappa$-deformed algebra, this system
has a non-Hermitian Hamiltonian.
Then, here we show that we obtain a non-Hermitian Hamiltonian from first
principles by employing the $\kappa$-deformed algebra.
Moreover, from mapping this system onto the Jaynes-Cummings (JC) and
anti-Jaynes-Cummings (AJC) models, this allows us to propose the
$\kappa$-JC and $\kappa$-AJC models, respectively.
Our derivations are general and can be applied to other similar cases.

The remainder of this paper is organized as follows.
In Sec. \ref{sec:2ddo} we revise the solution of the (2+1) Dirac
oscillator and the mapping of this system onto the JC and AJC systems.
In Sec. \ref{sec:kJC} we propose the $\kappa$-JC and $\kappa$-AJC
models.
In Sec. \ref{sec:symmetries} we study the symmetries of the
$\kappa$-(A)JC Hamiltonian.
In Sec. \ref{sec:dynamics} the dynamics of the $\kappa$-JC model is
presented.
Finally, our conclusions are presented in Sec. \ref{sec:conclusion}.

\section{The Dirac Oscillator and the mapping onto
  the JC and AJC systems}
\label{sec:2ddo}

In this section, we briefly review the Dirac oscillator and the exact
mapping onto the JC and AJC systems.
The Dirac oscillator, first proposed by It\^o \etal
\cite{NCA.51.1119.1967} and then further developed by Moshinsky
\etal \cite{JPA.22.817.1989}, has been a usual model for
studying physical properties of systems in various branches of physics.
In the non-relativistic limit, the Dirac oscillator reduces to the
simple harmonic oscillator with strong spin-orbit coupling.
It was shown that the Dirac oscillator can be regarded as describing a
neutral particle interaction with a static linear electric field
\cite{EJP.16.135.1995}.
Recently, the one-dimensional Dirac oscillator has had its first
experimental realization \cite{PRL.111.170405.2013}, and it
also was proposed as a tabletop experiment for direct observation of the
corresponding analog of virtual pair creation on quantum measurement
backaction \cite{PRL.121.110401.2018}.
These results have  made the system more attractive from the point of
view of applications.
For a detailed approach to the Dirac oscillator see
Refs. \cite{Book.1998.Strange,Book.1996.Moshinsky}.

The Dirac oscillator is obtained by means of the nonminimal
coupling \cite{JPA.22.817.1989}
\begin{equation}
  \label{eq:coupling}
  \mathbf{p}\to\mathbf{p} \pm im\omega\beta\mathbf{r},
\end{equation}
in the Dirac equation, with $\mathbf{p}$ the momentum operator, $m$
the mass, $\omega$ the oscillator frequency, $\mathbf{r}$  the
position vector, and $\beta$ a Dirac matrix.
The double signal introduced in Eq. \eqref{eq:coupling}
leads us to similar results \cite{ACP.1334.249.2011} and serves to
map the Dirac oscillator onto the JC (AJC) model for $+$ ($-$) in a
transparent manner.
The Dirac oscillator in (2+1) dimensions, when the third
spatial coordinate is absent, was studied in
Refs. \cite{EPL.108.30003.2014,PRA.77.033832.2008,PRA.76.041801R.2007,
PRA.49.586.1994}.
This system is achieved by writing the Dirac equation in (2+1)
dimensions including the nonminimal interaction in
Eq. \eqref{eq:coupling},
\begin{equation}
  \label{eq:dirac21}
  H^{\pm}\ket{\psi} =
  \left(
    c \boldsymbol{\alpha}\cdot\boldsymbol{\pi}^{\pm}+\beta m c^2
  \right)
  \ket{\psi} = E\ket{\psi},
\end{equation}
where $\ket{\psi}$ is a two-component spinor,
$\boldsymbol{\alpha}=\beta\boldsymbol{\gamma}$,
$\boldsymbol{\pi}^{\pm}=\mathbf{p} \pm im\omega\beta\mathbf{r}$
and the $2 \times 2$ Dirac matrices are defined in terms of the Pauli
matrices
\cite{PRL.64.503.1990}
\begin{equation}
  \label{eq:mset}
  \beta = \gamma_{0}=\sigma_{z},\qquad
  \beta\gamma_{1} =  \sigma_{x}, \qquad
  \beta\gamma_{2} = s\sigma_{y}.
\end{equation}
The parameter $s$ is twice the spin value and here serves to
characterize the two possible chiralities of the system, with
$s=-1$ ($s=+1$) corresponding to the left (right) chirality.
The approach employed here based on the matrix set \eqref{eq:mset}
differs from the usual one which chooses one specific value of the
chirality $s$ and has the advantage of making the results dependent on
the chirality in a transparent manner.
Thus, considering the two-component spinor as
$\ket{\psi}=(\ket{\psi_{1}},\ket{\psi_{2}})^T$, from
Eq. \eqref{eq:dirac21} we arrive at the following set of coupled
equations:
\begin{align}
  \label{eq:coupledset}
  (E-mc^{2})\ket{\psi_{1}} = {}
  &
    c (\pi_{x}^{\mp} - i s \pi_{y}^{\mp})\ket{\psi_{2}},
    \nonumber\\
  (E+mc^{2})\ket{\psi_{2}} = {}
  & c (\pi_{x}^{\pm} + i s \pi_{y}^{\pm})\ket{\psi_{1}},
\end{align}
where $\pi_{i}^{\pm}=p_i \pm i m \omega r_{i}$, $i=x,y$.
Introducing the chiral creation and annihilation operators
\cite{PRA.76.041801R.2007}
\begin{align}
  a_{s}^{\pm}= {}
  &
    \frac{1}{\sqrt{2}}(a_{x}^{\pm}
    \mp isa_{y}^{\pm}),
\end{align}
where  $a_{i}^{+}$ ($a_{i}^{-}$) is the usual creation
(annihilation)  operators of the usual harmonic oscillator,
\begin{equation}
  \label{eq:asho}
  a^{\pm}_{i}=\frac{1}{\sqrt{2}}\left(\frac{1}{\Delta}r_{i}
    \mp i\frac{\Delta}{\hbar}p_{i}\right),
\end{equation}
and $\Delta=\sqrt{\hbar/m\omega}$ is the ground state oscillator width,
Eqs. \eqref{eq:coupledset} can be written as
\begin{align}
  \label{eq:coupledspinor}
  (E-mc^{2})\ket{\psi_{1}} = {}
  & 2imc^{2}\sqrt{\xi}a^{\mp}_{s} \ket{\psi_{2}},
    \nonumber\\
  (E+mc^{2})\ket{\psi_{2}} = {}
  & -2imc^{2}\sqrt{\xi}a_{s}^{\pm}\ket{\psi_{1}},
\end{align}
with $\xi=\hbar\omega/mc^{2}$ representing the relativistic parameter
which leads to the nonrelativistic limit when $\xi \to 0$.
By squaring \eqref{eq:coupledspinor}, we find
\begin{align}
  \label{eq:decoupledspinor}
  (E^2-m^2c^{4})\ket{\psi_{1}} = {}
  &
    4m^2c^{4}\xi a^{\mp}_{s} a_{s}^{\pm} \ket{\psi_{1}},
    \nonumber\\
  (E^2-m^2c^{4})\ket{\psi_{2}} = {}
  & 4m^2c^{4}\xi a_{s}^{\pm}a_{s}^{\mp}\ket{\psi_{2}}.
\end{align}
Introducing the chiral quanta basis
\begin{equation}
  \label{eq:chiralket}
  \ket{n_{s}^{\pm}} =
  \frac{1}{\sqrt{n_{s}^{\pm}!}}(a_{s}^{+})^{n_{s}^{\pm}}\ket{0},
\end{equation}
with $n_{s}^{+}= 0,1,2,\ldots$ and  $n_{s}^{-}= 1,2,3\ldots$
representing the eigenvalues of the number operator,
$N_{s}=a_{s}^{+}a_{s}^{-}$, it is possible to
diagonalize both equations simultaneously.
In this manner, with $\ket{\psi_{1}}=\ket{n_{s}^{\pm}}$,
$\ket{\psi_{2}}=\ket{\tilde{n}^{\pm}_{s}}$ and due to the fact these
states represent the components of the same state vector with energy
$E^{\pm}$, we conclude that
$\tilde{n}^{\pm}_{s} = n_{s}^{\pm} \pm 1$,
and the energy eigenvalues are given by
\begin{equation}
  \label{eq:energyDO}
  E^{\pm}
  = \pm E_{n_{s}^{\pm}}^{\pm}
  = \pm mc^{2}\sqrt{1+4\xi \left[n_{s}^{\pm}+\Theta(\pm)\right]},
\end{equation}
where we have made use of the Heaviside step function
$\Theta(\pm)=(1 \pm 1)/2$.
We observe that the particle and antiparticle spectrum are symmetric
and, as we shall show shortly, the deformation breaks this symmetry.
These energy eigenvalues should be compared with those obtained by the
directed solution of the second-order differential equation in polar
coordinates that arises from the position representation of the Dirac
equation.
The result seems to be \cite{EPL.108.30003.2014}
\begin{equation}
  \label{eq:energyevppos}
  E^{\pm} = \pm E_{n}^{\pm}
  = \pm mc^{2}
  \sqrt{1+4\xi\left[n + (|l| - s l)/2 + \Theta(\pm)\right]},
\end{equation}
where $n=0,1,2,\ldots$ is the radial quantum number and
$l = 0, \pm 1, \pm 2, \ldots$ is the angular momentum quantum number.
So, the comparison leads to $n_s^{\pm}= n +(|l|-sl)/2$, showing the
dependency on $s$ and the high degeneracy of the (2+1) Dirac oscillator
spectra \cite{PLB.738.44.2014}.

The Hamiltonian  $H^{\pm}$ for the (2+1) Dirac oscillator can be
rewritten as
\begin{equation}
  \label{eq:HDO}
  H^{\pm}=
  \begin{pmatrix}
    mc^{2} &
    \mp 2 i m c^2 \sqrt{\xi}a_{\pm s}^{\mp}\\
    \pm 2 i m c^2 \sqrt{\xi} a_{\pm s}^{ \pm} & -mc^{2}
  \end{pmatrix}.
\end{equation}
As shown in \cite{PRA.76.041801R.2007}, using the notation
$\sigma_{z}=\ketbra{e}-\ketbra{g}$, $\sigma^{+}=\ketbra{e}{g}$, and
$\sigma^{-}=\ketbra{g}{e}$, in which $\sigma^{\pm}$ are the standard
fermionic two-level transition operators that obey the commutation
relation $[\sigma^{+},\sigma^{-}]=\sigma_{z}$ and $\ket{g}$ and
$\ket{e}$ are, respectively, the ground and excited states of a two
level quantum system, the Hamiltonian $H^{+}$ can be mapped onto the JC
model of quantum optics,
\begin{align}
  \label{eq:hplus}
  H^{+}
  =  {} &
          2 i m c^2 \sqrt{\xi}
          \left(
          a_{s}^{-}\ketbra{e}{g}
         -a_{s}^{+}\ketbra{g}{e}
          \right)
          +mc^2\sigma_{z}
          \nonumber \\
  = {} &
         \hbar\left(
         g a_{s}^{+}\sigma^{-}
         +
         g^{*} a_{s}^{-}\sigma^{+}
         \right)
         +\delta\sigma_{z}
         \nonumber,\\
  = {} & H_{\text{JC}}^{s},
\end{align}
where $g=2imc^2\sqrt{\xi}/\hbar$ is the coupling constant and
$\delta=mc^{2}$  is the detuning parameter proportional to the rest mass.
In an analogous manner, the AJC model can be obtained from $H^{-}$,
\begin{align}
  \label{eq:hminus}
  H^{-} = {}
  &
    \hbar
    \left(
    g a_{-s}^{+}\sigma^{+}+
    g^{*}a_{-s}^{-}\sigma^{-}
    \right)
    +\delta\sigma_{z}
    \nonumber\\
  = {}
  &
    H_{\text{AJC}}^{-s}.
\end{align}
Thus, the mapping onto the JC or AJC systems may be accomplished by a
suitable choice of the  nonminimal coupling signal in
Eq. \eqref{eq:coupling}, which amounts to the substitution
$\omega \to -\omega$.
Besides that, the substitution of the oscillator frequency turns
the JC system into the AJC system with opposite chirality, which is
evident when comparing \eqref{eq:hplus} with \eqref{eq:hminus}.
The results presented here are generalizations of results present
in the literature.
Thus using the double signal in the nonminimal coupling together with
the $s$ parameter, the mapping of the Dirac oscillator onto the JC and
AJC models is now more transparent.

\section{The $\kappa$-JC  and the  $\kappa$-AJC models}
\label{sec:kJC}

In this section, we present the $\kappa$-deformed Dirac oscillator, and
using the mapping of the previous section, we propose the
$\kappa$-deformed JC and AJC models.
The deformation studied here differs from previous models proposed in
the literature in the sense that it arises naturally from the
$\kappa$-deformed algebra.
It is interesting to comment that there are other proposals in the
literature for deformed (A)JC models, namely the q-deformed
\cite{EPL.120.44002.2017} and the f-deformed \cite{EPJP.132.458.2017}
models.
Both models are based on the deformation of the commutation relations
for the creation and annihilation operators and lead to Hermitian
Hamiltonians.
In the scenario presented here, the deformation stems from the
$\kappa$-deformed algebra and does not affect the creation and
annihilation operators and leads naturally to a non-Hermitian
Hamiltonian.

The $\kappa$-deformed Dirac equation in (2+1) dimensions
can be written as \cite{PLB.359.339.1995}
\begin{multline}
  \label{eq:diracdef}
  \big\{\gamma_{0}P_{0}-c\gamma_{i}P_{i}
  +\frac{\epsilon}{m c^2}\left[\gamma_{0}(P^{2}_{0}-P_{i}P_{i})
    -mc^{2}P_{0}\right]\big\}\ket{\psi} \\
  = mc^2\ket{\psi},
\end{multline}
where $\epsilon = m c^2 \varepsilon/2$ is the dimensionless deformation
parameter.
To obtain the $\kappa$-Dirac oscillator equation, we can proceed by
gauging Eq. \eqref{eq:diracdef} introducing the nonminimal coupling
of the previous section.
Thus gauging the above equation with the nonminimal coupling
prescription in Eq. \eqref{eq:coupling},
\begin{align}
  \label{eq:gauge}
P_{0}& \to P_{0} = H = E,\\
P_{i}& \to \pi_{i}^{\pm}=p_{i} \pm im\omega\beta r_{i},
\end{align}
we can write Eq. \eqref{eq:diracdef} as
\begin{align}
  \label{eq:gauged}
  H \ket{\psi}
  = {} &
         \Big[
         \left(c\gamma_0\gamma_{i}\pi_{i}^{\pm}+\gamma_0 m c^2\right)
         \nonumber \\
         &
         -\frac{\epsilon}{m c^2}
         \left(
         H^2- \pi_{i}^{\pm} \pi_{i}^{\pm}-\gamma_{0} mc^2 H
         \right)\Big] \ket{\psi}.
\end{align}
In general, noncommutative Hamiltonians should be addressed by
employing the Seiberg-Witten transformation, as discussed in
\cite{EPJP.134.332.2019}.
However, the gauge in Eq. \eqref{eq:gauge} leads to a commutative
Hamiltonian and, consequently, we do not need to deal with the
Seiberg-Witten transformation here.
Nevertheless, the above equation is quite complicated to solve
without using some sort of approximation.
A common approach \cite{PLB.318.613.1993} to solve it is to recognize
the first term in parentheses as the undeformed Hamiltonian [see
Eq. \eqref{eq:dirac21}], and iterate it only keeping terms up to
$O(\epsilon)$, leading to
\begin{equation}
  \label{eq:kappaDO}
  H _{\epsilon}^{\pm} \ket{\psi} = E \ket{\psi},
\end{equation}
with
\begin{equation}
  \label{eq:iterated}
  H_{\epsilon}^{\pm}=
  H^{\pm}
  -\frac{\epsilon}{m c^2}
  \left[
  \left(H^{\pm}\right)^2- \pi_{i}^{\pm} \pi_{i}^{\pm}
  -\gamma_0 mc^2H^{\pm}
  \right].
\end{equation}
Equation \eqref{eq:kappaDO} defines the (2+1) $\kappa$-Dirac oscillator
\cite{PLB.738.44.2014}.

\begin{figure}[t!]
  \centering
  \includegraphics[width=0.65\columnwidth]{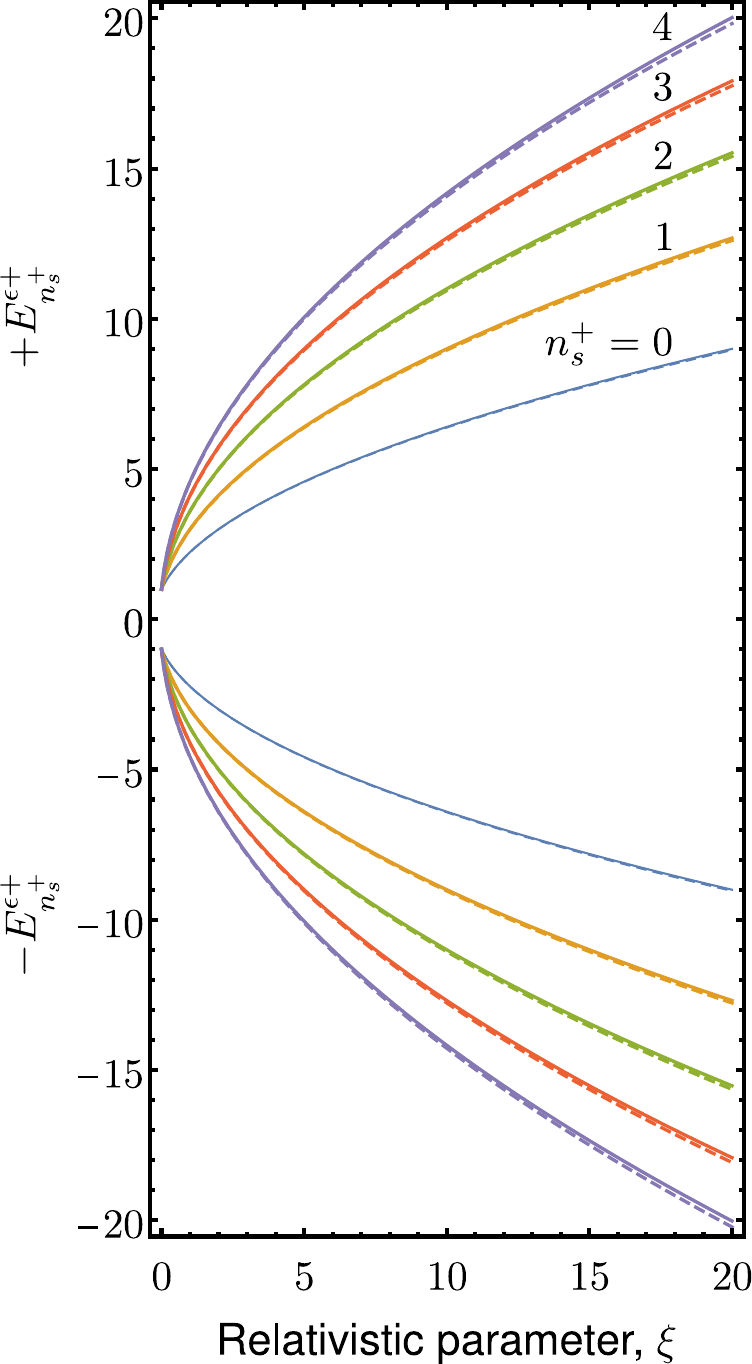}
  \caption{(Color online)
    JC (solid lines) and the $\kappa$-JC (dashed lines) eigenenergies
    $E_{n_s}^{\epsilon+}$ as a function of the relativistic parameter
    $\xi$, in units such as $\hbar=m=c=1$, for $n_s^{+}=0,\ldots,4$.
    The value used for the dimensionless deformation parameter is
    $\epsilon=5\times 10^{-4}$.
  }
  \label{fig:figa}
\end{figure}

We now proceed by employing the same reasoning used in the previous
section.
Thus using the representation of the $\gamma$ matrices as in Eq.
\eqref{eq:mset}, considering a two-component spinor and also introducing
the chiral creation and annihilation operators, the deformed Hamiltonian
$H_{\epsilon}^{\pm}$  can be written as
\begin{align}
  \label{eq:kJC}
  H_{\epsilon}^{\pm} = {}
  &
    \hbar \left(g\hat{a}^{+}_{\pm s}\sigma^{\mp}
    \mu_{\epsilon}^{\mp}
    +
    g^{*}\hat{a}^{-}_{\pm s}\sigma^{\pm}
    \mu_{\epsilon}^{\pm}\right)
    +\delta_{\epsilon}^{\pm}\sigma_z
    \nonumber \\
  &
    -2 mc^{2}\epsilon\xi(2\hat{N}_{\pm s}+1) \mathbbm{1},
\end{align}
with $\delta_{\epsilon}^{\pm}=(1 \mp 2\epsilon\xi)\delta$,
$\mu_{\epsilon}^{\pm}=1\pm \epsilon$, and $\mathbbm{1}$ the
identity matrix.
Notice that for $\epsilon=0$ we get back the Hamiltonians
in Eqs. \eqref{eq:hplus} and \eqref{eq:hminus}, \ie, the JC or AJC
models, respectively.
In this manner, by the mapping of the previous section, we propose the
Hamiltonian in Eq. \eqref{eq:kJC} as representing the $\kappa$-JC and
$\kappa$-AJC models:
\begin{equation}
  H_{\kappa\text{--JC}}^{s} = H_{\epsilon}^{+}, \qquad
  H_{\kappa\text{--AJC}}^{-s} = H_{\epsilon}^{-}.
\end{equation}
It is important to note that due to the presence of
$\mu_{\epsilon}^{\pm}$ in  $H_{\epsilon}^{\pm}$, which comes from the
term $\epsilon\gamma_{0}mc^2H^{\pm}$ in the deformed Hamiltonian in
Eq. \eqref{eq:iterated}, it fails to be Hermitian, \ie,
$H_{\epsilon}^{\pm}\neq {H_{\epsilon}^{\pm}}^{\dagger}$.
As a consequence, $H_{\epsilon}^{\pm}$ being non-Hermitian leads to a
non-unitary time evolution.
Nevertheless, the spectrum of the allowed energy eigenvalues of
$H_{\epsilon}^{\pm}$ is real, and is given by
\begin{equation}
  \label{eq:EnergykJC}
  E_{n_s^{\pm}}^{\epsilon\pm}
  = \pm E_{n_s^{\pm}}^{\pm}
  - 4 m c^2 \xi \epsilon [n_{s}^{\pm}+\Theta(\pm)],
\end{equation}
which coincides with the result obtained in \cite{PLB.738.44.2014} and
immediately reduces to the undeformed energy eigenvalues  in
Eq. \eqref{eq:energyDO} for $\epsilon=0$.
We observe that the deformation causes an asymmetric energy shift in the
energy eigenvalues when compared with the standard (A)JC model,
$|\Delta^{\epsilon} E|=4mc^2\xi\epsilon[n_{s}^{\pm}+\Theta(\pm)]$,
which increases with $\xi$ and is larger for larger values of
$n_s^{\pm}$.
In the context of the $\kappa$-Dirac oscillator, this asymmetry stems
from the fact that the deformed Hamiltonian breaks the charge
conjugation symmetry \cite{PLB.731.327.2014,PLB.738.44.2014}.
The energy spectrum of the $\kappa$-(A)JC has a positive energy branch
which is bounded from below by $mc^2\sqrt{1+4\xi}-4mc^2\xi\epsilon$ and
a negative branch bounded from above by
$-mc^2\sqrt{1+4\xi}-4mc^2\xi\epsilon$, as displayed in
Fig. \ref{fig:figa} for the JC and $\kappa$-JC.
The graph for the AJC and  $\kappa$-AJC looks identical.
Following Ref. \cite{PLB.738.44.2014}, in which an upper bound for the
deformation parameter was obtained, we used value
$\epsilon=5 \times 10^{-4}$ as the value for the dimensionless
deformation parameter.

\section{Symmetries and the non-Hermiticity of the
  $\kappa$-JC and $\kappa$-AJC models }
\label{sec:symmetries}

As we stated above, $H_{\epsilon}^{\pm}$ is non-Hermitian and it
leads to a non-unitary evolution.
In fact, the $\kappa$-deformed Hamiltonian is not even
$\mathcal{PT}$-symmetric, but it is quasi-Hermitian.
We can check this by first looking at the effects of parity and time
reversal symmetry operations on $a$'s and $\sigma$'s operators
\cite{Book.2011.Sakurai}:
\begin{align*}
  \mathcal{P} a_{s}^{\pm} \mathcal{P}^{-1} &= a_{-s}^{\pm},
  &
  \mathcal{P} \sigma_{z} \mathcal{P}^{-1}&=\sigma_{z},
  &
   \mathcal{P} \sigma^{\pm} \mathcal{P}^{-1}&=\sigma^{\pm},\\
  \mathcal{T} a_{s}^{\pm} \mathcal{T}^{-1} &= -a_{s}^{\pm},
  &
  \mathcal{T}\sigma_{z} \mathcal{T}^{-1} &= -\sigma_{z},
  &
   \mathcal{T} \sigma^{\pm} \mathcal{T}^{-1} &= \sigma^{\mp}.
 \end{align*}
Through this, one notes that the deformed Hamiltonian in
Eq. \eqref{eq:kJC} is not $\mathcal{PT}$-symmetric since
\begin{align}
  \label{eq:kJC-PT}
  \mathcal{PT}H_{\epsilon}^{\pm}(\mathcal{PT})^{-1} = {}
  &
    \hbar
    \left(
    g  a^{+}_{\pm s}\sigma^{\mp}
    \mu_{\epsilon}^{\mp}
    +
    g^{*} a^{-}_{\pm s}\sigma^{\pm}
    \mu_{\epsilon}^{\pm}
    \right)
    - \delta_{\epsilon}^{\pm}\sigma_z
    \nonumber \\
  &
    -2 mc^{2}\epsilon\xi(2N_{\pm s}+1) \mathbbm{1}
    \nonumber \\
  \neq  {}
  &
    H_{\epsilon}^{\pm}.
\end{align}
However, the Hamiltonian is invariant under the transformation
$\mathcal{P}\sigma_{z}$, so that
\begin{align}
  \label{eq:kJC-sigmaz}
  \mathcal{P}\sigma_zH_{\epsilon}^{\pm}(\mathcal{P}\sigma_z)^{-1}
  = H_{\epsilon}^{\pm}.
\end{align}
This symmetry was also observed in a similar system in
Ref. \cite{MPLA.20.655.2005}.
Another interesting transformation is given by
\begin{align}
  \label{eq:kJC-Tx}
  \mathcal{T}\sigma_{x}H_{\epsilon}^{\pm}
  {(\mathcal{T}\sigma_{x})}^{-1}
  = {}
  &
    \hbar \left(g a^{+}_{\mp s}\sigma^{\mp}
    \mu_{\epsilon}^{\mp}
    +
    g^{*} a^{-}_{\mp s}\sigma^{\pm}
    \mu_{\epsilon}^{\pm}\right) + \delta_{\epsilon}^{\pm}\sigma_z
    \nonumber \\
  &
    -2 mc^{2}\epsilon\xi(2N_{\mp s}+1) \mathbbm{1},
\end{align}
which leads to the original Hamiltonian but with the chirality changed,
$s\rightarrow -s$.
Although the Hamiltonian is not $\mathcal{PT}$-symmetric, it is
quasi-Hermitian since its eigenvalues are real \cite{PRA.91.052113.2015}.
Therefore, the Hamiltonian satisfies a quasi-Hermiticity relation
\begin{align}
\label{eq:qherm}
  {H_{\epsilon}^{\pm}}^{\dagger} =
  \eta_{\pm}H_{\epsilon}^{\pm} \eta_{\pm}^{-1},
\end{align}
for some positive-defined operator
$\eta_{\pm}$ \footnote{In order to restrict to theories with
positive-definite norm}, the so-called metric operator, which defines
the inner-product
\begin{align}
  \label{eq:inner-product}
  \langle \cdot,\cdot \rangle_{\eta_{\pm}} =
  \langle \cdot,\eta_{\pm}\cdot \rangle,
\end{align}
with respect to which the Hamiltonian is said to be Hermitian since
\begin{align}
  \langle \phi,H_{\epsilon}^{\pm}\psi
  \rangle_{\eta_{\pm}}
  = {}
  &
    \langle \phi \vert \eta_{\pm}
    H_{\epsilon}^{\pm} \vert \psi \rangle \nonumber \\
  = {}
  &
    \langle \phi \vert {H_{\epsilon}^{\pm}}^{\dagger}
    \eta_{\pm} \vert \psi \rangle \nonumber \\
  = {}
  &
  \langle {H_{\epsilon}^{\pm}}\phi,\psi
    \rangle_{\eta_{\pm}},
\end{align}
for all $\phi$ and $\psi$ in the domain of $H_{\epsilon}^{\pm}$.
Thus, decomposing the metric operator as
$\eta_{\pm}= {\rho_{\pm}}^{\dagger}\rho_{\pm}$,
Eq. \eqref{eq:qherm} allows us to define a Hermitian counterpart
associated with the non-Hermitian one,
\begin{align}
  h_{\epsilon}^{\pm} =
  \rho_{\pm}H_{\epsilon}^{\pm}{\rho_{\pm}}^{\dagger},
\end{align}
in such a way that
$h_{\epsilon}^{\pm}={h_{\epsilon}^{\pm}}^{\dagger}$.
The expected values in both representations are the same
\begin{align}
  \langle H_{\epsilon}^{\pm}\rangle_{\eta_{\pm},\Phi}
  = {}
  &
    \langle \Phi \vert \eta_{\pm}H_{\epsilon}^{\pm} \vert
    \Phi \rangle \nonumber \\
  = {}
  &
    \langle \Phi \vert
    {\rho_{\pm}}^{\dagger}h_{\epsilon}^{\pm}\rho_{\pm}
    \vert \Phi \rangle \nonumber \\
  = {}
  &
    \langle \Psi \vert h_{\epsilon}^{\pm} \vert \Psi
    \rangle \nonumber \\
  = {}
  &
    \langle h_{\epsilon}^{\pm}\rangle_{\Psi},
\end{align}
with $\vert \Psi \rangle = \rho_{\pm} \vert \Phi \rangle$.

\section{System dynamics}
\label{sec:dynamics}

Until now we have worked with both $\kappa$-JC and $\kappa$-AJC
systems simultaneously.
For the sake of clarity, in what follows we focus our discussion on the
$\kappa$-JC Hamiltonian, $H^{s}_{\kappa\text{--JC}}=H_{\epsilon}^{+}$.
At the end we comment briefly on how to obtain the results for
the $\kappa$-AJC Hamiltonian,
$H^{-s}_{\kappa\text{--AJC}}=H_{\epsilon}^{-}$.
Thus, to simplify the notation, we drop the $+$ signal in our
equations.
To obtain the Hermitian operator associated with
$H_{\kappa\text{--JC}}^{s}$, a suitable similarity transformation is
given by the operator
\begin{align}
  \label{eq:rhomapJC}
  \rho=
  e^{\epsilon a^{-}_{s}a^{-}_{s}-\epsilon a^{+}_{s} a^{+}_{s}+\epsilon
  a^{+}_{s} a^{-}_{s}},
\end{align}
satisfying $\rho^{\dagger} \rho=\eta$.
Note that this operator is a function of creation and annihilation
operators with same chiralities and reduces to the identity operator for
$\epsilon=0$.
Thus, the Hermitian operator can be obtained from the similarity
transformation
\begin{align}
  h_{\kappa \text{--JC}}^{s} = \rho H_{\kappa\text{--JC}}^{s}\rho^{-1},
\end{align}
with $\rho^{\dagger}\rho=\eta$.
The result seems to be
\begin{align}
  h^{s}_{\kappa\text{--JC}}
  = {}
  &
    \hbar(g \sigma^{-}a_{s}^{+}+
    g^{*}\sigma^{+}a_{s}^{-})
    +(1 - 2\epsilon\xi)\delta\sigma_z
    \nonumber\\
  & +
    \epsilon\hbar
    (g\sigma^{-}a_{s}^{-}+g^{*}\sigma^{+}a_{s}^{+})
    \nonumber \\
  &
    -2 mc^{2}\epsilon\xi(2N_{s}+1) \mathbbm{1}.
\end{align}
The spectrum of the Hermitian operator $h_{\kappa\text{--JC}}^{s}$ is given
by  \eqref{eq:EnergykJC} as it shares the spectrum with the
non-Hermitian operator $H_{\kappa\text{--JC}}^{s}$.
To find the associated deformed energy eigenstates for the positive
and negative deformed energy eigenvalues of $h_{\kappa\text{JC}}^{s}$, we
first solve the eigenvalue equation for the non-Hermitian operator,
$H_{\kappa\text{--JC}}^{s}\ket{E_{n_s}^{\epsilon}} =
E_{n_s}^{\epsilon}\ket{E_{n_s}^{\epsilon}}$, then apply the
transformation $\rho$.
In this manner, using the Pauli spinors
$\ket{{\uparrow}}=(1,0)^{\dagger}$ and
$\ket{{\downarrow}}=(0,1)^{\dagger}$, the deformed energy
eigenstates of $H_{\kappa\text{--JC}}^{s}$ can be written as
\begin{align}
  \label{eq:StateskJC}
  \ket{\pm E_{n_s}^{\epsilon}} = {}
  &
  \sqrt{\frac{E_{n_s}\pm mc^2}{2E_{n_s}}}
    \ket{n_{s}}\ket{{\uparrow}} \nonumber \\
  &
  \pm
  i \sqrt{\frac{E_{n_s}\mp mc^2}{2E_{n_s}}} (1-\epsilon)
  \ket{n_{s}+1}\ket{{\downarrow}},
\end{align}
and by applying the transformation $\rho$, we obtain
\begin{align}
  \label{eq:stateskJCh+}
  \ket{+\bar{E}_{n_{s}}^{\epsilon}}
  = {}
  & \rho \ket{+E_{n_{s}}^{\epsilon}}
    \nonumber\\
  = {}
  &
    \alpha_{n_{s}}  \ket{n_{s}}\ket{{\uparrow}}
    +i \beta_{n_{s}} \ket{n_{s}+1}\ket{{\downarrow}}
    \nonumber\\
  &
    + \epsilon \alpha_{n_{s}}
    \left(
    c_{n_s}   \ket{n_{s}-2}\ket{{\uparrow}}
    -c_{n_{s}+2}\ket{n_{s}+2}\ket{{\uparrow}}
    \right)
    \nonumber\\
  &
    +i\epsilon \beta_{n_{s}}
    \left(
    c_{n_{s}+1} \ket{n_{s}-1}\ket{{\downarrow}}
    -c_{n_{s}+3}\ket{n_{s}+3}\ket{{\downarrow}}
    \right),
\end{align}
and
\begin{align}
  \label{eq:stateskJCh-}
  \ket{-\bar{E}_{n_{s}}^{\epsilon}}
  = {}
  & \rho \ket{-E_{n_{s}}^{\epsilon}}
    \nonumber\\
  = {}
  &
    \beta_{n_{s}}  \ket{n_{s}}\ket{{\uparrow}}
    -i\alpha_{n_{s}} \ket{n_{s}+1}\ket{{\downarrow}}
    \nonumber\\
  &
    +\epsilon \beta_{n_{s}}
    \left(
    c_{n_{s}}  \ket{n_{s}-2}\ket{{\uparrow}}
    -c_{n_{s}+2} \ket{n_{s}+2}\ket{{\uparrow}}
    \right)
    \nonumber\\
  &
    -i\epsilon\alpha_{n_{s}}
    \left(
    c_{n_{s}}   \ket{n_{s}-1}\ket{{\downarrow}}
    -c_{n_{s}+3}\ket{n_{s}+3}\ket{{\downarrow}}
    \right),
\end{align}
with
\begin{equation}
  \alpha_{n_{s}}=
  \sqrt{\dfrac{E_{n_s}+mc^{2}}{2 E_{n_s}}},
  \qquad
  \beta_{n_{s}}=
  \sqrt{\dfrac{E_{n_s}-mc^{2}}{2 E_{n_s}}},
\end{equation}
and $c_{n_s}=\sqrt{n_s(n_s-1)}$.
Note that the eigenstates in \eqref{eq:stateskJCh+} and
\eqref{eq:stateskJCh-} are normalized up to first order in $\epsilon$
and, as observed in \cite{PRA.76.041801R.2007}, these eigenstates show
that the spin and angular momentum are entangled.
Moreover, the presence of deformation gives rise to new entangled
states.
Equations \eqref{eq:stateskJCh+} and  \eqref{eq:stateskJCh-} allow us to
write an initial state
$\ket{\Psi_{n_s}(0)} = \ket{n_{s}}\ket{{\uparrow}}$ in terms  of the
positive- and negative-energy eigenstates, namely
\begin{align}
  \label{eq:inistate}
  \ket{\Psi_{n_s}(0)}  = {}
  &
    \alpha_{n_{s}}\ket{+\bar{E}_{n_{s}}^{\epsilon}}
    +\beta_{n_{s}}\ket{-\bar{E}_{n_{s}}^{\epsilon}}
    \nonumber \\
  &
    -\epsilon c_{n_{s}}
    \left(
    \alpha_{n_{s}-2}\ket{+\bar{E}_{n_{s}-2}^{\epsilon}}
    +\beta_{n_{s}-2}\ket{-\bar{E}_{n_{s}-2}^{\epsilon}}
    \right)
    \nonumber \\
  &
    +\epsilon c_{n_{s}+2}
    \left(
    \alpha_{n_{s}+2}\ket{+\bar{E}_{n_{s}+2}^{\epsilon}}
    +\beta_{n_{s}+2}\ket{-\bar{E}_{n_{s}+2}^{\epsilon}}
    \right).
\end{align}
This superposition of the positive- and negative-energy eigenstates is a
signature of the \textit{Zitterbewegung}, which here is
encoded in the spin degree of freedom, and can be associated to Rabi
oscillations due to the interference of these eigenstates
\cite{PRA.76.041801R.2007}.
The \textit{Zitterbewegung} is a relativistic quantum effect generally
understood as a trembling motion of relativistic particles
\cite{Book.2000.Greiner}, difficult to be measured, but can be simulated
experimentally in one dimension \cite{N.463.68.2010}.
We observe again that the deformation introduces more eigenstates in the
superposition and for $\epsilon=0$ our results immediately reduce to
the ones in Ref. \cite{PRA.76.041801R.2007}.

Now that we have a Hermitian operator and its eigenstates, we can
proceed to study the system dynamics.
Thus starting with \eqref{eq:inistate}, it leads to a state at time $t$
given by
\begin{widetext}
\begin{align}
  \ket{\Psi_{n_s}(t)} = {}
  &
    \alpha_{n_{s}}\ket{+\bar{E}_{n_{s}}^{\epsilon}}
    e^{-i\omega^{\epsilon +}_{n_{s}}t}
    +\beta_{n_{s}}\ket{-\bar{E}_{n_{s}}^{\epsilon}}
    e^{-i\omega^{\epsilon -}_{n_{s}}t}
    \nonumber\\
  &
    - \epsilon c_{n_s}
    \alpha_{n_{s}-2}\ket{+\bar{E}_{n_{s}-2}^{\epsilon}}
    e^{-i\omega^{\epsilon +}_{n_{s}-2}t}
    - \epsilon c_{n_s}
     \beta_{n_{s}-2}\ket{-\bar{E}_{n_{s}-2}^{\epsilon}}
     e^{-i\omega^{\epsilon -}_{n_{s}-2}t}
  \nonumber \\
  &
    +\epsilon c_{n_s+2}
    \alpha_{n_{s}+2}\ket{+\bar{E}_{n_{s}+2}^{\epsilon}}
    e^{-i\omega^{\epsilon +}_{n_{s}+2}t}
    +\epsilon c_{n_s+2}
     \beta_{n_{s}+2}\ket{-\bar{E}_{n_{s}+2}^{\epsilon}}
     e^{-i\omega^{\epsilon -}_{n_{s}+2}t},
\end{align}
where
\begin{equation}
  \omega^{\epsilon
    \pm}_{n_{s}}=\pm\omega_{n_{s}}-\phi_{n_{s}}^{\epsilon},
\end{equation}
is the $\kappa$-deformed \textit{Zitterbewegung} frequency, with
$\omega_{n_{s}}=E^{+}_{n_{s}}/\hbar$, and
$\phi_{n_{s}}^{\epsilon} = 4m^{2}c^{4}\epsilon\xi(n_{s}+1)/\hbar$.
Now, writing the evolved state in the language of Pauli spinors, we have
\begin{align}
  \label{eq:Eq.Ev.States}
  \ket{\Psi_{n_s}(t)} = {}
  &
    e^{i\phi_{n_{s}}^{\epsilon}t}
    \left(
    f_{n_{s}}(t)\ket{n_{s}}\ket{{\uparrow}}
    +g_{n_{s}}(t)\ket{n_{s}+1}\ket{{\downarrow}}
    \right)
    \nonumber\\
  &
    +\epsilon e^{i\phi_{n_{s}}^{\epsilon}t}
    \left[
    c_{n_{s}}f_{n_{s}}(t) \ket{n_{s}-2}\ket{{\uparrow}}
    +c_{n_{s}+1}g_{n_{s}}(t) \ket{n_{s}-1}\ket{{\downarrow}}
    \right]
  \nonumber\\
  &
    -\epsilon e^{i\phi_{n_{s}}^{\epsilon}t}
    \left(
    c_{n_{s}+2} f_{n_{s}}(t) \ket{n_{s}+2}\ket{{\uparrow}}
    -c_{n_{s}+3}g_{n_{s}}(t)\ket{n_{s}+3}\ket{{\downarrow}}
    \right)
  \nonumber\\
  &
    -\epsilon c_{n_{s}} e^{i\phi_{n_{s}-2}^{\epsilon}t}
    \left(
    f_{n_{s}-2}(t)\ket{n_{s}-2}\ket{{\uparrow}}
    +g_{n_{s}-2}(t)\ket{n_{s}-1}\ket{{\downarrow}}
    \right)
  \nonumber\\
  &
    +\epsilon c_{n_{s}+2} e^{i\phi_{n_{s}+2}^{\epsilon}t}
    \left(
    f_{n_{s}+2}(t)\ket{n_{s}+2}\ket{{\uparrow}}
    +g_{n_{s}+2}(t)\ket{n_{s}+3}\ket{{\downarrow}}
    \right),
\end{align}
\end{widetext}
where
\begin{equation}
  f_{n_{s}}(t) = \cos(\omega_{n_{s}}t)
  -\frac{i\sin(\omega_{n_{s}}t)}{\sqrt{1+4\xi(n_{s}+1)}},
\end{equation}
and
\begin{equation}
  g_{n_{s}}(t)=2\sin(\omega_{n_{s}}t)\alpha_{n_{s}}\beta_{n_{s}},
\end{equation}
with $|f_{n_{s}}(t)|^{2}+|g_{n_{s}}(t)|^{2}=1$.

We can appreciate the modifications caused by the $\kappa$-deformation
by evaluating the expectation values of the $z$ component of the spin,
orbital and total angular momentum observables, which are defined by
\begin{equation}
  S_z = \frac{\hbar}{2}\sigma_z,
  \quad
  L_z=\hbar (N_s-N_{-s}),
  \quad
  J_z=L_z+S_z,
\end{equation}
respectively.
Surprisingly, even though the $\kappa$-deformation modifies the energy
eigenvalues, gives rise to new entangled states with different quantum
numbers, and modifies the \textit{Zitterbewegung} frequency,
\textit{there is no first-order correction on the expectation values
of the $\kappa$-JC}.
This kind of result was already observed in the $\kappa$-Dirac-Coulomb
problem \cite{PLB.318.613.1993}, where the first-order correction on
this system is identically zero.

\begin{figure}[t]
  \centering
  \includegraphics[width=\columnwidth]{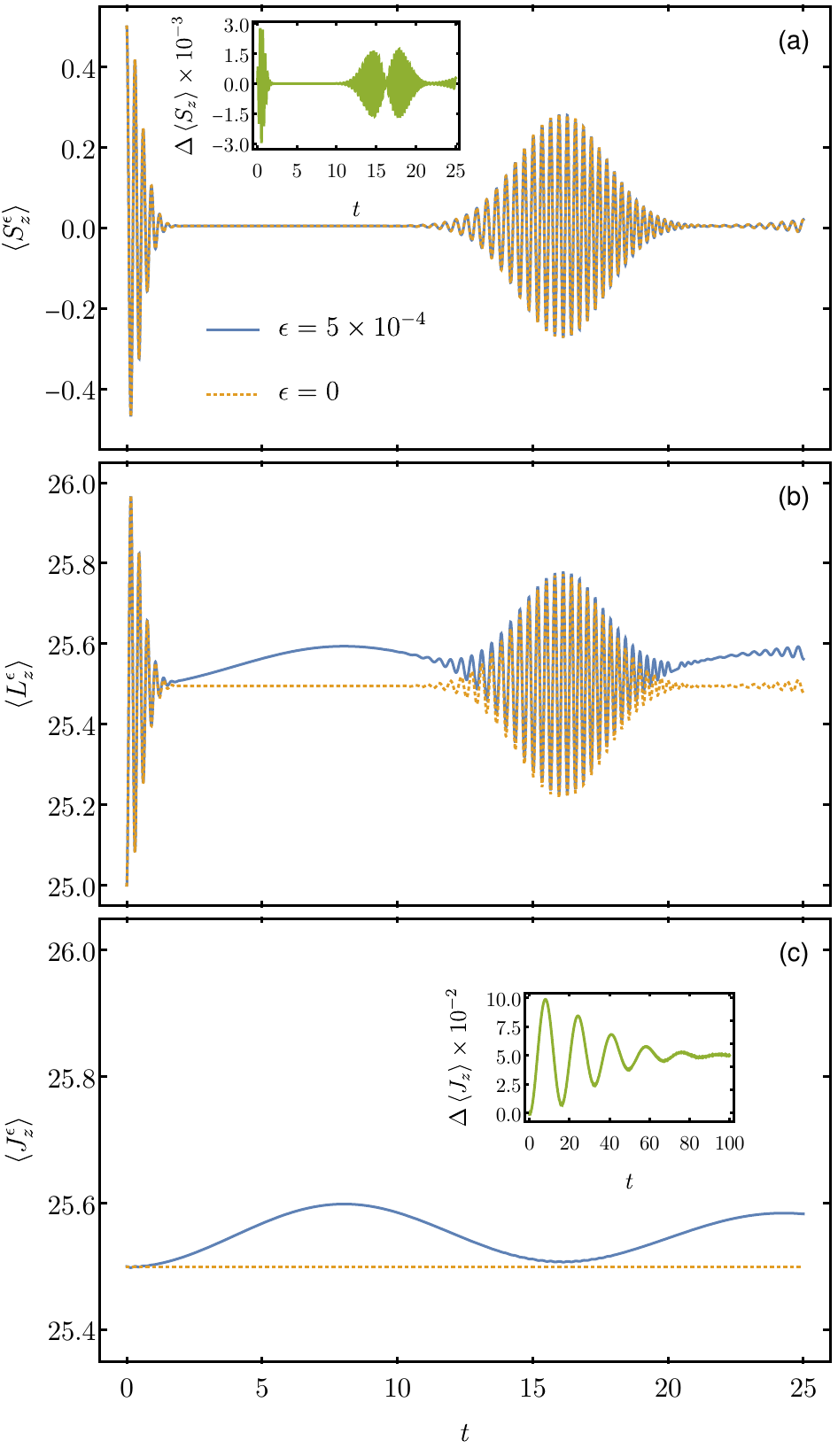}
  \caption{(Color online)
    Behavior of the expectation values
    (a) $\expval{S_z^{\epsilon}}$,
    (b) $\expval{L_z^{\epsilon}}$, and
    (c) $\expval{J_z^{\epsilon}}$ as a function of time for a system
    with mean photon number $\expval{n_s}=25$ for
    $\epsilon=5\times10^{-4}$ (blue solid line) and
    $\epsilon=0$ (orange dotted line) using units such as
    $m=\hbar=\omega=c=1$.
    In (a), the two curves are superposed and the inset shows the
    difference between them.
    In (c), the inset shows the asymptotic behavior of the
    $\Delta\expval{J_z^{\epsilon}}$.
  }
  \label{fig:fig2}
\end{figure}

On the other hand, we can observe first-order effects of the
$\kappa$-deformation on the scenario of collapsed and revivals of
the atomic population in the $\kappa$-JC model by employing an initial
coherent state.
Thus, considering the initial state as
$\ket{\Psi_{n_s}(0)}=\ket{\alpha}\ket{{\uparrow}}$, with
\begin{equation}
  \ket{\alpha} =
  e^{-\frac{|\alpha|^{2}}{2}}
  \sum_{n_{s}=0}^{\infty}\frac{\alpha^{n_{s}}}{\sqrt{n_{s}!}}\ket{n_{s}},
\end{equation}
the $\kappa$-deformed expectation values are given by
\begin{align}
  \expval{S_{z}^{\epsilon}} = {}
  &
    \frac{\hbar}{2}
    -\hbar \sum_{n_{s}=0}^{\infty}
    \frac
    {\expval{n_s}^{n_{s}}e^{-\expval{n_s}}}
    {n_{s}!} \mathcal{S}_{n_s}(t)
    \nonumber\\
  &
    + \hbar \epsilon \sum_{n_{s}=0}^{\infty}
    \frac{\expval{n_s}^{n_{s}+1}e^{-\expval{n_s}}}{n!}
    [\mathcal{S}_{n_s}(t)-\mathcal{S}_{n_s+2}(t)],
\end{align}
\begin{align}
  \expval{L_{z}^{\epsilon}} = {}
  &
    \hbar\expval{n_s}
    +\hbar\sum_{n_{s}=0}^{\infty}
    \frac
    {\expval{n_s}^{n_{s}}e^{-\expval{n_s}}}
    {n_{s}!} \mathcal{S}_{n_s}(t)
    \nonumber \\
  &
    -\hbar \epsilon \sum_{n_{s}=0}^{\infty}
    \frac{\expval{n_s}^{n_{s}+1}e^{-\expval{n_s}}}{n!}
    [\mathcal{S}_{n_s}(t)-\mathcal{S}_{n_s+2}(t)]
    \nonumber \\
  &
    +\hbar\epsilon
   \sum_{n_{s}=0}^{\infty}
    \frac{\expval{n_s}^{n_{s}+1} e^{-\expval{n_s}}}{n!}
    \mathcal{L}_{n_{s}}^{\epsilon}(t),
\end{align}
and
\begin{align}
  \expval{J_{z}^{\epsilon}} = {}
  &
    \hbar\left[\expval{n_s}+\frac{1}{2}\right]
    +\hbar\epsilon
   \sum_{n_{s}=0}^{\infty}
    \frac{\expval{n_s}^{n_{s}+1} e^{-\expval{n_s}}}{n!}
    \mathcal{L}_{n_{s}}^{\epsilon}(t),
\end{align}
where $\expval{n_s}=|\alpha|^2$,
\begin{align}
  \mathcal{S}_{n_s}(t) = {}
  &
    \frac{4 \xi (n_{s}+1)}{[1+4\xi (n_{s}+1)]}
    \sin^{2}{(\omega_{n_s}t)},
\end{align}
\begin{align}
  \mathcal{L}_{n_s}^{\epsilon}(t) = {}
  &
    \frac{4c_{n_s+2}}{\sqrt{(n_{s}+2)!}}
    - \frac{2c_{n_s+2}}{\sqrt{(n_{s}+2)!}}
    w_{n_{s}}(t)\cos(\Phi^{\epsilon}t)
    \nonumber \\
  &
    - \frac{2c_{n_s+2}}{\sqrt{(n_{s}+2)!}}
    s_{n_{s}}(t)\sin(\Phi^{\epsilon}t)
    \nonumber \\
  &
    - \frac{2c_{n_s+3}}{\sqrt{(n_{s}+2)!}}
    p_{n_{s}}(t)\cos(\Phi^{\epsilon}t),
\end{align}
with
\begin{equation}
  \Phi^{\epsilon}
  = \phi_{n_{s}}^{\epsilon}-\phi_{n_{s}+2}^{\epsilon}
  = -8 m^2c^4 \xi \epsilon,
\end{equation}
\begin{align}
  w_{n_{s}}(t) = {}
  &
    2\cos(\omega_{n_{s}}t)\cos(\omega_{n_{s}+2}t)
    \nonumber \\
  &
  +\frac
  {2\sin(\omega_{n_{s}}t)\sin(\omega_{n_{s}+2}t)}
  {\sqrt{[1+4\xi(n_{s}+1)][1+4\xi(n_{s}+3)]}},
\end{align}
\begin{align}
  s_{n_{s}}(t) = {}
  &
    \frac{2\cos(\omega_{n_{s}+2}t)\sin(\omega_{n_{s}}t)}
    {\sqrt{1+4\xi(n_{s}+1)}}
    \nonumber \\
  &
    -\frac{2\cos(\omega_{n_{s}}t)\sin(\omega_{n_{s}+2}t)}
    {\sqrt{1+4\xi(n_{s}+3)}},
\end{align}
and
\begin{equation}
  p_{n_{s}}(t)=
  2\alpha_{n_{s}}\alpha_{n_{s}+2}\sin(\omega_{n_{s}}t)\sin(\omega_{n_{s}+2}t).
\end{equation}
We can observe that the deformation modifies all the expectation values.
To help us analyze the effects of the deformation on the expectation
values, let us define
\begin{equation}
  \Delta\expval{O}=\expval{O^{\epsilon}}-\expval{O},
\end{equation}
as the difference between the $\kappa$-deformed expectation value of the
observable $O$ and the usual (undeformed) one.
Figure \ref{fig:fig2} shows the results for the expectation values
as a function of time $t$  for a system with mean photon number
$\expval{n_s}=25$, using units such as $m=\hbar=\omega=c=1$ and
$\epsilon=5 \times 10^{-4}$ (blue solid lines)
and for $\epsilon=0$ (orange dotted lines).
In Fig. \ref{fig:fig2}(a) we show $\expval{S_z^{\epsilon}}$ and
it shows the well-known initial collapse followed by the revival
of the spin inversion.
The inset shows $\Delta \expval{S_z}$ and we observe that the
expectation value is slightly modified by the deformation.
On the other hand, in Fig. \ref{fig:fig2}(b) we show the
$\expval{L_{z}^{\epsilon}}$ and we can also observe collapse and
revival, but now the orbital angular momentum is noticeably more
affected by the deformation than the spin angular momentum.
As a result, we observe that $\expval{J_{z}^{\epsilon}}$ is not
constant of motion anymore when the deformation is present,
as we can see in Fig. \ref{fig:fig2}(c).
So, the $\kappa$-deformed expectation value of the $z$ component of the
total angular momentum is not a conserved quantity.
This result can be understood by noting that $J_z$ fails to commute
with $h_{\kappa\text{--JC}}^{s}$,
\begin{equation}
  [h_{\kappa\text{--JC}}^{s},J_{z}] =
  4 \epsilon \hbar g
  (\sigma^{-}\hat{a}^{-}_{s}+\sigma^{+}\hat{a}^{+}_{s}),
\end{equation}
and this failure is a direct consequence of the deformation.
We can also observe that deformation displaces the expectation value of
the $J_z$ [see the inset in Fig \ref{fig:fig2}(c)] and for large values
of $t$ it converges to a fixed amount,
$\Delta \expval{J_{z}}|_{t \to \infty} \sim 5\times 10^{-2}$.
It is easy to see that for $\epsilon=0$, $J_z$ commutes with the
Hamiltonian and we recover all the results of the usual JC system, as it
should be.

Finally, we end up saying that for the $\kappa$-AJC Hamiltonian, we
observe that a suitable similarity transformation is
\begin{align}
  \label{eq:rhomapAJC}
  \rho=
  e^{\epsilon a^{-}_{s}a^{-}_{s}-\epsilon a^{+}_{s} a^{+}_{s}-\epsilon
  a^{+}_{s} a^{-}_{s}},
\end{align}
in which, in comparison with the map for the $\kappa$-JC in Eq. \eqref{eq:rhomapJC}, the last term has a sign reversal, and leads us
to the following Hermitian $\kappa$-AJC Hamiltonian
\begin{align}
  h_{\kappa \text{--AJC}}^{-s} = {}
  &
    \rho H_{\kappa\text{--AJC}}^{-s}\rho^{-1}
  \nonumber \\
  = {}
  &
    \hbar(g \sigma^{+}a_{-s}^{+}+
    g^{*}\sigma^{-}a_{-s}^{-})
    +(1 + 2\epsilon\xi)\delta\sigma_z
    \nonumber\\
  & +
    \epsilon\hbar
    (g\sigma^{+}a_{-s}^{-}+g^{*}\sigma^{-}a_{-s}^{+})
    \nonumber \\
  & -2
    mc^{2}\epsilon\xi(2N_{-s}+1) \mathbbm{1}.
\end{align}
Thus it is straightforward to show that similar results can be obtained
for the $\kappa$-AJC system by using the Hermitian Hamiltonian
$h_{\kappa \text{--AJC}}^{-s}$.

\section{Conclusion}
\label{sec:conclusion}

In conclusion, we have revisited the Dirac oscillator in (2+1)
dimensions and its mapping onto the JC and AJC models.
The mapping is now transparent as we have made the connection
between the non-minimal coupling signal $+$ ($-$) of the Dirac
oscillator Hamiltonian and the JC (AJC) model.
We have also introduced the parameter $s=\pm 1$ to characterize the two
possible chiralities, allowing one to discuss them simultaneously.
By considering the (2+1) Dirac oscillator in the context of the
$\kappa$-deformed algebra and using the above mapping, we have proposed
the $\kappa$-JC and the $\kappa$-AJC models.
We have shown that the $\kappa$-deformation leads naturally to a
non-Hermitian Hamiltonian, something that leads to a non-unitary time
evolution.
Moreover, the $\kappa$-(A)JC Hamiltonian is not even
$\mathcal{PT}$-symmetric, but is quasi-Hermitian as it possesses a real
spectrum, and by employing the theory of quasi-Hermitian Hamiltonians,
we have found its Hermitian counterpart, allowing us to study the
dynamics of the $\kappa$-deformed system.
Although the displacement was caused by the deformation on the
eigenenergies and, consequently, on the \textit{Zitterbewegung}
frequencies, we have observed no first-order effects on the expectation
values of $S_z$, $L_z$, and $J_z$, when considering an initial state such
as $\ket{\Psi_{n_s}(0)}=\ket{n_s}\ket{\uparrow}$.
On the other hand, when considering a coherent initial state, we have
observed modifications on the well-known collapse and revival behavior,
as well as on the above expectation values.
Especially, we have observed that the expectation value of the total
angular momentum in the $z$ direction, $J_z$, is not a constant of
motion anymore as a direct consequence of the $\kappa$-deformation.

We comment that the mapping between quantum optical and relativistic
quantum systems \cite{JPA.32.5367.1999} led to a great breakthrough
in quantum simulation experiments of relativistic quantum effects
\cite{N.463.68.2010} since direct measurements of relativistic
quantum phenomena are not easy to do.
Significant examples of relativistic quantum effects simulated
through optical setups are the experimental simulation of the
\textit{Zitterbewegung} effect in trapped ions \cite{PRL.98.253005.2007}
and others \cite{PRL.94.206801.2005,PRL.85.4643.2000}.
As shown in \cite{PRA.76.041801R.2007}, the dynamics of the 2D
Dirac oscillator can be implemented in a single trapped ion inside a
Paul trap, and given the fact these systems allow a vast coherent
control of ionic internal and external degrees of freedom
\cite{RMP.75.281.2003}, and the ability to tune experimental parameters
that could also introduce certain modifications that would entail novel
phenomena, our work suggests that some future experiment might be able
to detect the effects of the $\kappa$-deformation presented here.

\section*{Acknowledgments}
The authors are grateful to Professor M. Moussa for helpful discussions.
This work was partially supported by the Brazilian agencies Conselho
Nacional de Desenvolvimento Cient\'ifico e Te\-cnol\'ogico (CNPq),
Funda\c{c}\~{a}o Arauc\'{a}ria (FAPPR, Grant No. 09/2016) and Instituto
Nacional de Ci\^{e}ncia e Tecnologia de Informa\c{c}\~{a}o Qu\^{a}ntica
(INCT-IQ).
It was also financed by the Co\-or\-dena\c{c}\~{a}o de
Aperfei\c{c}oamento de Pessoal de N\'{i}vel Superior (CAPES, Finance
Code 001).
F.M.A. also acknowledges CNPq Grants No. 434134/2018-0 and
No. 314594/2020-5.

%

\end{document}